\documentstyle[emulateapj]{article}

\newcommand{\cha}{{\sl Chandra}}
\newcommand{\ein}{{\sl Einstein}}
\newcommand{\hub}{{\sl Hubble}}
\newcommand{\ros}{{\sl ROSAT}}

\slugcomment{Submitted to Astrophysical Journal Letters.}

\begin{document}

\title{
DISCOVERY OF SPATIAL AND SPECTRAL STRUCTURE IN THE X-RAY EMISSION FROM THE CRAB NEBULA}

\author{
Martin C. Weisskopf\altaffilmark{1}, 
J. Jeff Hester\altaffilmark{2},
Allyn F. Tennant\altaffilmark{1}, 
Ronald F. Elsner\altaffilmark{1},\\
Norbert S. Schulz\altaffilmark{3},
Herman L. Marshall\altaffilmark{3}, 
Margarita Karovska\altaffilmark{4}, 
Joy S. Nichols\altaffilmark{4},\\ 
Douglas A. Swartz\altaffilmark{5},
Jeffery J. Kolodziejczak\altaffilmark{1}, and
Stephen L. O'Dell\altaffilmark{1}
} 

\altaffiltext{1}{Space Science Department, NASA Marshall Space Flight Center, NASA/MSFC/SD50, Huntsville, AL 35812.} 
\altaffiltext{2}{Department of Physics and Astronomy, Arizona State University, Tyler Mall, Tempe, AZ 85287.} 
\altaffiltext{3}{Center for Space Research, Massachusetts Institute of Technology, Cambridge, MA 02139.} 
\altaffiltext{4}{Harvard--Smithsonian Center for Astrophysics, 60 Garden Street, Cambridge, MA 02138.}
\altaffiltext{5}{Universities Space Research Association, NASA Marshall Space Flight Center, USRA/MSFC/SD50, Huntsville, AL 35812.}

\begin{abstract}
The \cha\ {\sl X-ray Observatory} observed the Crab Nebula and Pulsar during orbital calibration.
Zeroth-order images with the High-Energy Transmission Grating (HETG) read-out by the Advanced CCD Imaging Spectrometer spectroscopy array (ACIS-S) show a striking richness of x-ray structure, at a resolution comparable to that of the best ground-based visible-light observations.
The HETG--ACIS-S images reveal, for the first time, an x-ray inner ring within the x-ray torus, the suggestion of a hollow-tube structure for the torus, and x-ray knots along the inner ring and (perhaps) along the inward extension of the x-ray jet.
Although complicated by instrumental effects and the brightness of the Crab Nebula, the spectrometric analysis shows systematic variations of the x-ray spectrum throughout the Nebula. 
\end{abstract}


\section{Introduction} \label{s:introduction}

The Crab Nebula is the archetypal filled-center supernova remnant, or {\em plerion} (\cite{Weiler1978}). 
Unlike shell-like supernova remnants, a central pulsar presumably powers each filled-center supernova remnant.
Thus the inner nebula of a plerion is particularly interesting, because it is the site of conversion of pulsar-supplied energy into synchrotron-emitting electrons.
Although this general picture seems valid, details of the coupling mechanism (e.g., \cite{Rees1974}; \cite{Dobrowolny1976}; \cite{Kennel1984}; \cite{Michel1985}; \cite{Emmering1987}; \cite{Coroniti1990}; \cite{Begelman1992}; \cite{Michel1994}) remain uncertain.

Detailed studies (\cite{Hester1995}) with the \hub\ {\sl Space Telescope} (HST) illustrate the rich structure of the Crab Nebula's synchrotron emission, on scales down to $0.2\arcsec$.
Especially intriguing are the structures of the inner nebula near the Pulsar~--- wisps (\cite{Scargle1969}), knots, and fibrous texture (\cite{Hester1995})~--- which exhibit cylindrical symmetry.
\hub\ observations (\cite{Hester1998}) also show that wisps form and dissipate over a few weeks, while moving outward at about $0.5 c$, thus clarifying the dynamical nature of these features (\cite{Scargle1969}).

Even more dramatic are the x-ray images, which clearly show this cylindrical morphology.
In a remarkable synthesis from lunar-occultation, modulation-collimator, and low-resolution-imaging x-ray observations, Aschenbach and Brinkmann (1975) had interpreted the extant data in terms of an x-ray torus.
Subsequent, higher-resolution x-ray data~--- from \ein\ (\cite{Brinkmann1985}), \ros\ (\cite{Hester1995}; \cite{Greiveldinger1999}), and \cha\ (reported here)~--- confirm this morphology and detect a jet and counter jet nearly along the axis of the torus.

On 1999 July 23, NASA's shuttle Columbia launched the \cha\ {\sl X-Ray Observatory}. 
The half-power diameter of the \cha\ system-level point spread function is about $0.8\arcsec$, at least a factor-of-five better than that of any other x-ray observatory and comparable to that of the best visible-light ground observations.
Thus, \cha\ provides an excellent tool for probing the rich structure of the inner Crab Nebula in x rays.
Here, we report the first high-spatial-resolution observation (\S \ref{s:observation}) and describe the observed morphology (\S \ref{s:morphology}) and spectrum (\S \ref{s:spectrum}) of the Crab Nebula, as obtained with \cha.

\section{OBSERVATION AND ANALYSIS} \label{s:observation}

On 1999 August 29, during orbital calibration, \cha\ obtained a 2667-s observation of the Crab Nebula, using the High-Energy Transmission Grating (HETG, \cite{Canizares2000}) and the Advanced CCD Imaging Spectrometer spectroscopy array (ACIS-S, \cite{Garmire2000}) of 6 CCDs. 
The data include events in both zero and higher spectral order, the latter being dispersed along two crossed tracks, owing to the slightly different orientation of the High-Energy Grating (HEG) facets from the Medium-Energy Grating (MEG) facets.
Here, we examine only the zeroth-order image of the Nebula.

\placefigure{f:image-full}

Figure~\ref{f:image-full} shows the HETG--ACIS-S image of the Crab Nebula, for the centermost ACIS-S CCD~--- designated ``S3''~--- which is one of two back-illuminated CCDs in the 10-CCD ACIS focal plane.
The \cha\ X-ray Center's standard processing has applied aspect corrections and compensated for dither (which averages pixel-to-pixel variations of the detector) over a $16\arcsec \! \times \!  16\arcsec$ area.
Acquired in the {\em timed-exposure graded mode} at 3.241 s/frame, the data comprise an event list, with each event recorded in a $3 \! \times \!  3$-pixel ``island'', where each ACIS pixel is about $0.5\arcsec$ (actually, $0.492\arcsec$) square. 
In this mode, the on-board software detects events and grades them according to which pixels (including the central pixel) have enough charge to exceed pre-defined thresholds. 

Owing to the x-ray brilliance of the Crab Nebula and Pulsar, \cha\ observations of this source are particularly challenging.
First, the total count rate is so high that the ACIS data saturates telemetry, resulting in dropped ACIS frames. 
Although this results in lost events, it does so in an unbiased manner and is not a substantive problem.

\begin{figure} 
\resizebox{6in}{!}{\rotatebox{90}{\includegraphics{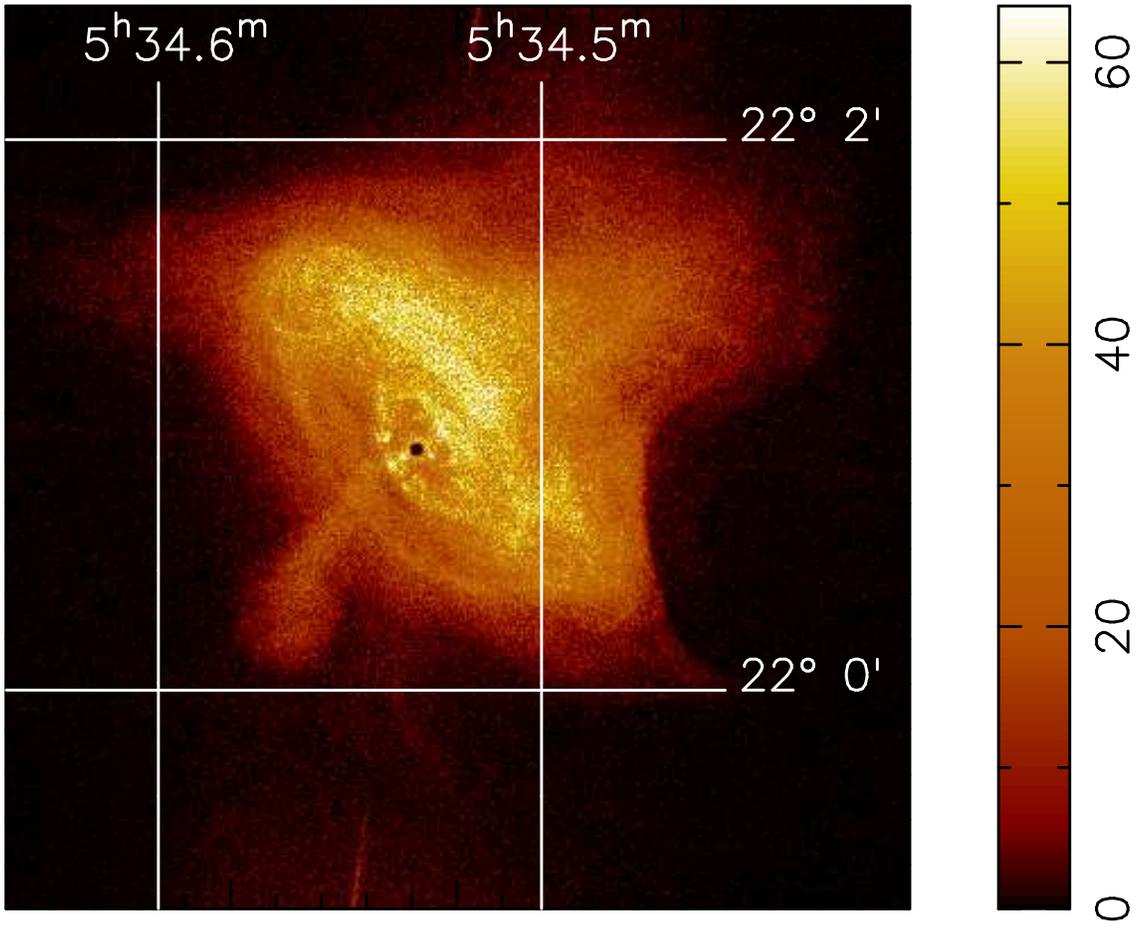}}}
\caption{\label{f:image-raw} Zoomed \cha\ HETG--ACIS-S image of the central $200\arcsec \! \times \!  200\arcsec$ of the Crab Nebula.
Previously undiscovered x-ray features include the inner ring and its knots, apparent lateral striations and circular structures at the toroid's extremities, and inward extension of the jet and counter jet.
The color bar indicates counts per pixel.
}
\end{figure}

More problematic is ``pile-up'', in which multiple photons contribute charge to the same or adjacent pixels, thus appearing as a single detected event.
First, pile-up effects ``grade migration'', changing the grade of a valid event to a grade which might normally be excluded.
To mitigate this, we retained all telemetered grades in the imaging analysis (\S \ref{s:morphology}), but not in the spectral analysis (\S \ref{s:spectrum}).
Second, in regions of high surface brightness, the detected event rate is not proportional to the photon rate; thus, reducing contrast.
In regions of very high surface brightness, the pile-up is so extreme that the charge deposited in a $3\! \times \! 3$-pixel island exceeds a pre-determined threshold (corresponding to about 15~keV), causing on-board rejection of the event.
This ``saturation'' results in no detected events in the very brightest areas of the image~--- namely, at the position of the Pulsar (Figures~\ref{f:image-full} and \ref{f:image-raw}).

To assess the impact of pile-up on the spectral distribution, we analyzed the ``trailed'' image (Figure~\ref{f:image-full}). 
The trailed image results from x rays which strike the CCD while the frame-transfer electronics is moving the exposed frame into the CCD's frame-store region. 
As each of the 1024 rows transfers toward the frame-store region, it spends about 40~$\mu s$ at each physical row.
Compared to the 3.20-s integration time per frame, the 41-ms frame-transfer time is quite small, so pile up is negligible for the trailed image.
Of course, the trailed image has no spatial information in the transfer (trail) direction.
 
The region from which we extracted the trailed image also contains photons from the timed exposure of the Nebula's x-ray halo. 
To subtract this component from the trailed image, we obtained a ``background'' from regions B (Figure~\ref{f:image-full}).  
A further complication is the presence in region B, of a component due to the trailed image of the first-order dispersed spectrum of the Nebula.
Because this contribution is small, we simply scaled the flux in region D of the dispersed image by $300\times (4\times 10^{-5})/3.20 = 0.0038$~--- number of rows occupied by the first-order image of the Nebula, times transfer time per row, divided by integration time per frame~--- and subtracted its spectrum from that of regions B.
This results in a 5\% decrease in the total rate of regions B, with most of the reduction occurring above 4~keV.
We then used this modified spectrum as the ``background'' in the trailed-image regions T.

\placefigure{f:image-raw}

\section{RESULTS --- MORPHOLOGY} \label{s:morphology}

Figure~\ref{f:image-raw} shows a $200\arcsec \! \times \!  200\arcsec$ image of the Crab Nebula, extracted from the S3 full-frame image (Figure~\ref{f:image-full}).
The \cha\ image clearly shows the x-ray torus (\cite{Aschenbach1975}) and jet and counterjet previously observed (\cite{Brinkmann1985}; \cite{Hester1995}; \cite{Greiveldinger1999}), which define the SE-to-NW axis of the inner Nebula.  
On somewhat larger scales, the image shows a sharply bounded notch (WSW of the Pulsar) into the x-ray nebular emission, earlier associated with the ``west bay'' of the Nebula (\cite{Hester1995}).   
Visible-light polarization maps of the Crab Nebula (\cite{Schmidt1979}; \cite{Hickson1990}) demonstrate that the magnetic field is parallel to the boundary of this notch, thus excluding the x-ray-emitting relativistic electrons from the west bay.

\begin{figure} 
\resizebox{6in}{!}{\rotatebox{90}{\includegraphics{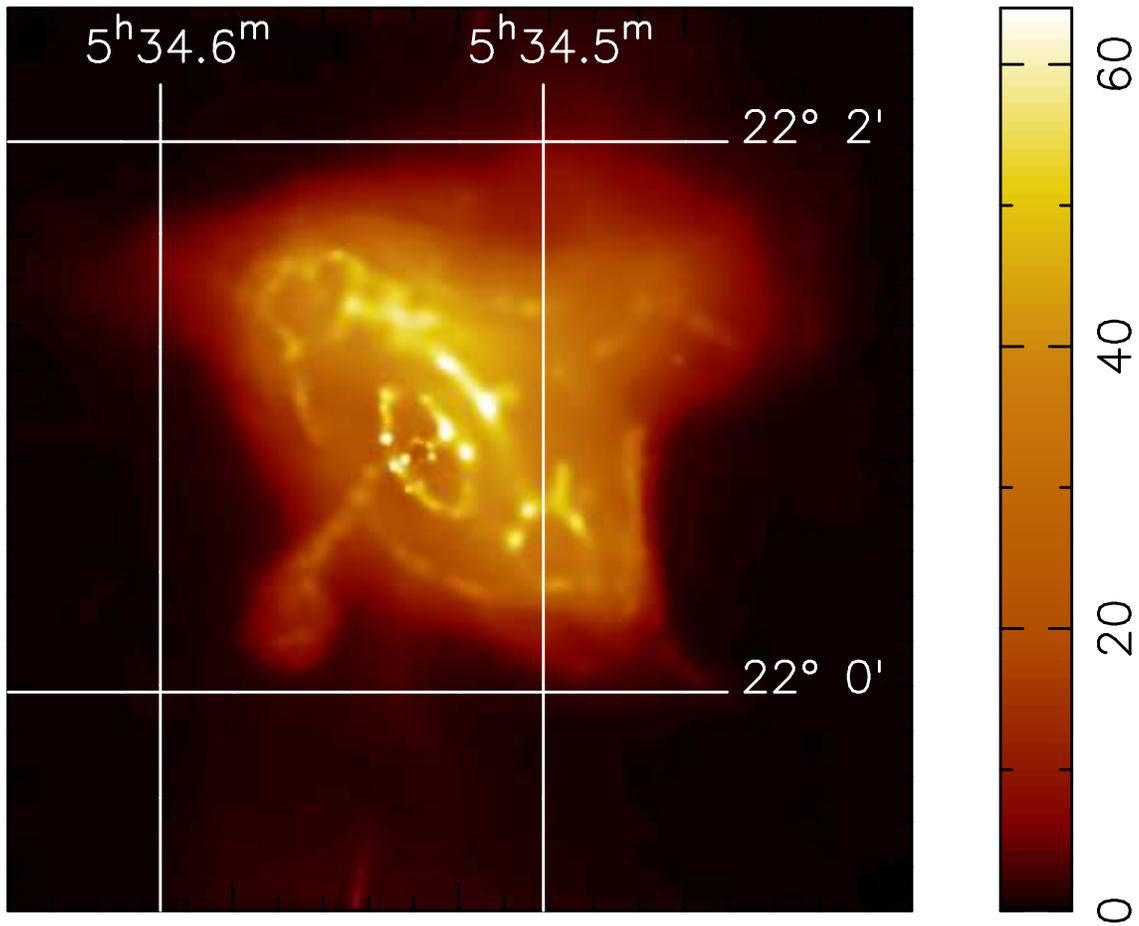}}}
\caption{\label{f:image-smooth} Adaptively smoothed \cha\ HETG--ACIS-S image of the central $200\arcsec \! \times \!  200\arcsec$ of the Crab Nebula.
The processed image accentuates small-scale structure, sometimes artificially; but is useful in identifying specific features in the raw image (Figure~\ref{f:image-raw}).
The color bar indicates counts per pixel.
}
\end{figure}

Of course, it is for scales from $1\arcsec$ to $10\arcsec$, that \cha\ provides unique information on x-ray structure.
To call attention to specific features, we processed the image (Figure~\ref{f:image-raw}) using the flux-conserving adaptive-smoothing algorithm CSMOOTH (\cite{Ebeling2000}), from the \cha\ Interactive Analysis of Observations (CIAO) package.
Figure~\ref{f:image-smooth} thus emphasizes small-scale structure, sometimes artificially.
To avoid misinterpretation, we use it only in conjunction with the raw image.

\placefigure{f:image-smooth}

The most striking feature of the x-ray image is the inner ring, lying between the Pulsar and the torus, which may correspond to a shock in the pulsar wind (\cite{Rees1974}; \cite{Kennel1984}).
Comparing the \hub\ image (\cite{Hester1995}) with the \cha\ image, we find that the positions of the visible-light ``wisp 1'' and the ``counter wisp'' correspond~--- at least in projection~--- approximately to the NW and SE quadrants, respectively, of the x-ray inner ring.
On the ring, reside a few compact (about $3\arcsec$) knots, one lying SE of the Pulsar, along the projected inward extension of the jet.
The surface brightness of this knot is too high to be explained as the superposition of the ring's and jet's surface brightnesses.

Having a semi-major axis of about $14\arcsec$ (0.14~pc) and a semi-minor axis of about $7\arcsec$ (0.07~pc), the projected rotational-symmetry axis of the inner ring lies about $58\arcdeg$ W of N.
The rotational-symmetry axis lies about $30\arcdeg$ out of the celestial plane, consistent with tilts estimated for the x-ray torus (\cite{Aschenbach1975}) and for various visible-light structures (\cite{Hester1995}). 

%% Based upon \cha\ observations (\cite{Helfand2000}) of the Vela Nebula, 
%% it appears that toroidal structures may be characteristic 
%% of plerionic supernova remnants.
%% The most remarkable feature of the Crab Nebula's x-ray torus 
%% is the circular structure at each extremity. 

The most provocative feature of the x-ray torus is the apparent circular structure at each extremity, which resemble the limb-brightened cross-section of a hollow-tube ring torus, with $8\arcsec$ (0.08~pc) tube radius.
Such structures, if indeed present, significantly constrain models for the inner Nebula.
The apparent tube geometry may indicate a role for relativistic ions with Larmor radii on this scale (\cite{Gallant1994}) or a toroidal current which generates a meridional magnetic field wrapping the torus.

Elsewhere, the torus exhibits circumferential fibrous texture but no knots.
As noted previously (\cite{Aschenbach1975}), the surface brightness varies significantly with toroidal azimuth, which has been interpreted (\cite{Pelling1987}; \cite{Greiveldinger1999}) as resulting from moderate relativistic beaming.
In contrast, the surface brightness of the inner ring is more uniform in azimuth (except for the knots), indicating that relativistic beaming is less significant for it than for the torus.
Measured from the center of the torus to the center of its tube, the semi-major and semi-minor axes are approximately $38\arcsec$ (0.37~pc) and $18\arcsec$ (0.17~pc), respectively.
The rotational-symmetry axis of the toroid lies about $48\arcdeg$ W of N and about $28\arcdeg$ out of the celestial plane.
Because the boundary of the torus is rather diffuse and dependent upon azimuth, these values are less certain than those for the inner ring.
Nevertheless, the in-plane orientation of the torus differs noticeably from that of the inner ring, suggesting that the torus is warped.

The \cha\ image probes the jet (to the SE) and counterjet (to the NW) at higher resolution than previous images with \ein\ (\cite{Brinkmann1985}) or with \ros\ (\cite{Hester1995}; \cite{Greiveldinger1999}).
Although the \cha\ image shows little additional structure or fibrous texture in the jet or counterjet (or ``cocoon'' encasing it), it traces them inward closer to the Pulsar than previously possible~--- at least to its projection onto the inner ring.
Besides the knot at the projected intersection of the jet with the inner ring, there seems to be another knot, between the inner ring and the Pulsar, on the extension of the jet to inside the inner ring.
Due to pile-up around the image of the Pulsar, we cannot exclude that this feature is an artifact.
On the other hand, \hub\ images (\cite{Hester1995}) exhibit similar structure (``knot~2'' and ``anvil'') in this region (a few arseconds SE of the Pulsar).
However, we need not expect features in the inner Nebula to be static (e.g., \cite{Scargle1969}; \cite{Hester1995}; \cite{Greiveldinger1999}).
Indeed, those regions showing the brightest x-ray emission tend to correspond to especially dynamic structure (\cite{Hester1998}), supporting the interpretation of these features as shocks in the polar jet (\cite{Hester1995}).

\section{RESULTS --- SPECTRAL VARIATIONS} \label{s:spectrum}

\begin{figure} 
\resizebox{6in}{!}{\rotatebox{90}{\includegraphics{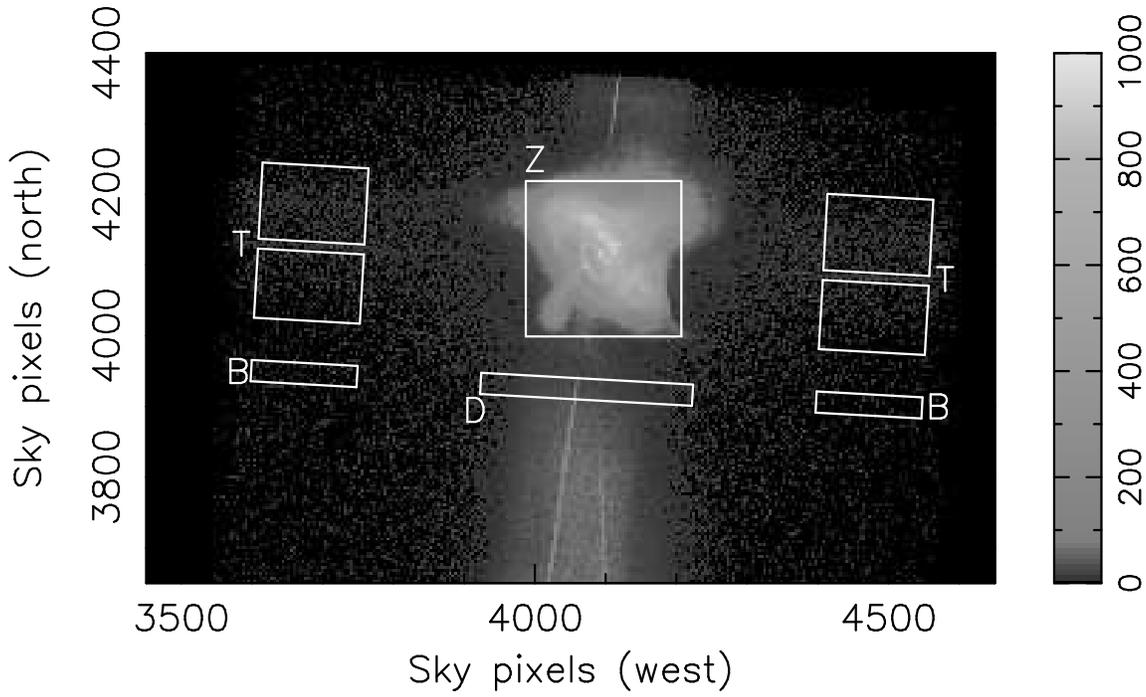}}}
\caption{\label{f:image-full} \cha\ HETG--ACIS-S image of the Crab Nebula, on CCD S3.
Each sky pixel is about $0.5\arcsec$ (actually, $0.492\arcsec$).
The nearly vertical, narrow X is the Pulsar's dispersed spectrum; the nearly horizontal line is its trailed image.
Boxes T denote regions used for analyzing the trailed image of the Nebula (\S \ref{s:spectrum}); boxes B, for determining halo surface brightness; and box D, for establishing the first-order trailed image in boxes B.
Box Z encloses the region examined for spectral-hardness variations (\S \ref{s:spectrum}) in the timed-exposure image.
The gray-scale bar indicates counts per $4 \! \times \!  4$-pixel spatial bin.
}
\end{figure}

In examining the spectral distribution, we use standard flight grades ({\sl ASCA} grades 02346), for which response matrices currently exist. 
First, we define the hardness $h_{2.7}$ as the ratio of flux above 2.7~keV to that below. 
This energy boundary fairly well separates events above the iridium-M$_V$ and gold-M$_V$ edges in the instrumental response from those below.
We establish from the trailed image (\S \ref{s:observation}), which is free of pile-up, that $h_{2.7} = 0.57 \pm 0.06$ for the Nebula as a whole.
To search for spectral variations within the Nebula, we determine $h_{2.7}$ for each $5\arcsec \! \times \!  5\arcsec$ bin of a $22 \! \times \!  22$ array (region Z of Figure~\ref{f:image-full}). 

\begin{figure} 
\resizebox{7in}{!}{\includegraphics{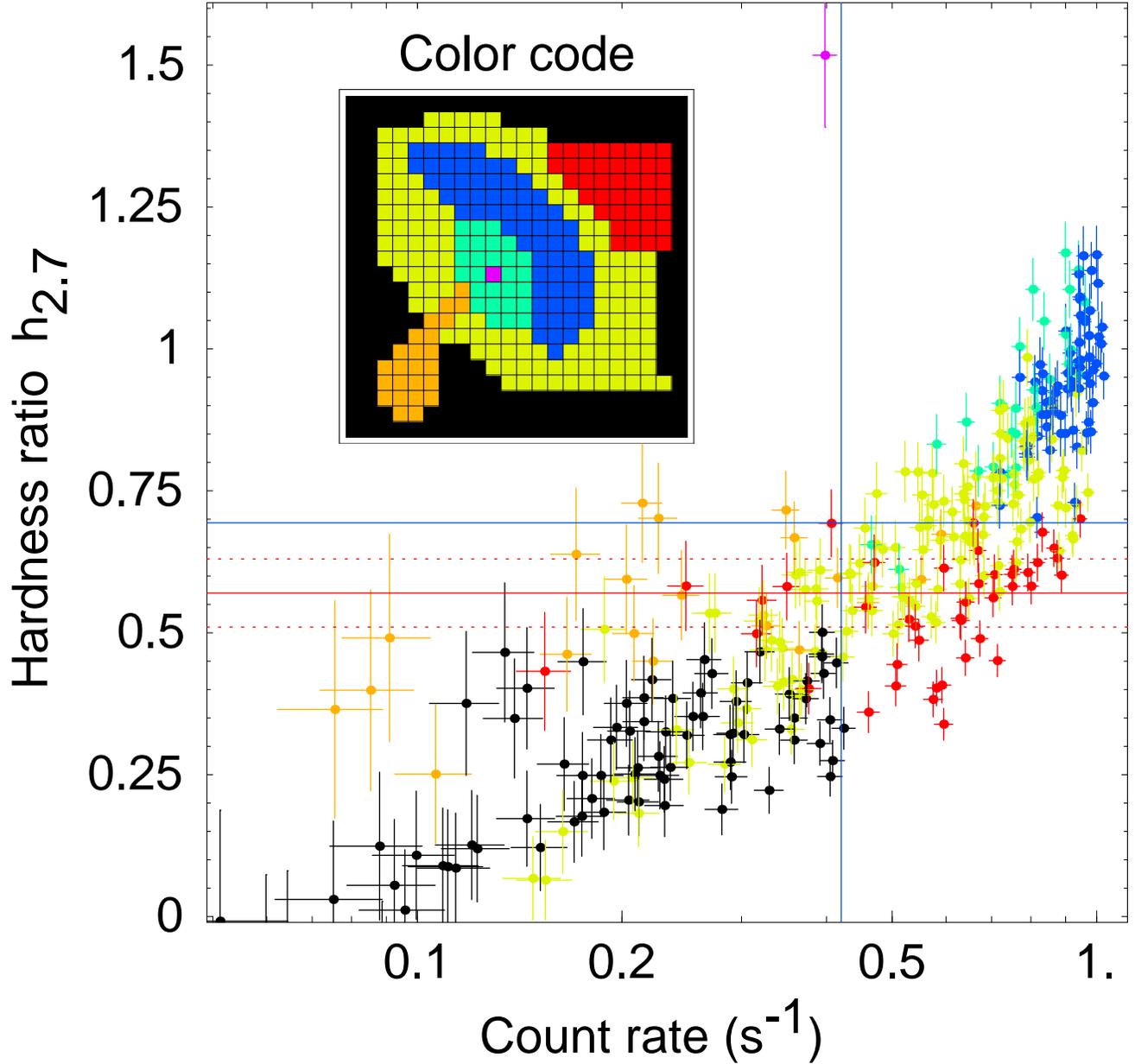}}
\caption{\label{f:hardness} Distribution of x-ray hardness in the Crab Nebula.
The scatter diagram, color-coded by location in the Nebula, shows hardness $h_{2.7}$ (\S \ref{s:spectrum}) against count rate in $5\arcsec \! \times \!  5\arcsec$ spatial bins.
The horizontal red line denotes average hardness (solid line) and associated error (dashed line) of the trailed image; the horizontal violet line, average hardness of the timed-exposure image, which is biased by pile-up effects (\S \ref{s:observation}).
The outer Nebula is softer than the trailed-image average; the jet and counter jet (and its cocoon) have about average hardness. 
Note that these data are not quantitatively correct at higher count rates, due to pile-up effects.}
\end{figure}

\placefigure{f:hardness}

Figure~\ref{f:hardness} displays the distribution of hardness $h_{2.7}$ within the Nebula.
For count rates below about 0.7~s$^{-1}$, many $5\arcsec \! \times \!  5\arcsec$ bins have hardness {\em less} than the nebular average.
This {\em cannot} result from pile-up, which only hardens the spectrum. Obviously, because there are many bins with hardness ratios {\em below} the average, some of the remaining bins must be {\em above} average. 
Thus, the majority of the high-count-rate data, although artificially harder due to pile-up, is nonetheless truly harder than the average~--- i.e., there are real variations of x-ray hardness within the Nebula.
The outer Nebula is significantly softer than the average, as expected for synchrotron losses and previously observed for the visible (\cite{Scargle1969}; \cite{Veron-Cetty1993}) and ultraviolet (\cite{Hennessy1992}) continua.
The jet and counter-jet (and its cocoon) have hardness comparable to the average; whereas, the inner ring appears to be harder.
However, the higher-contrast structure in the inner ring means that the peak surface brightness in each bin may significantly exceed the average in the bin, leading to more pile up than suggested by the mean surface brightness.
Consequently, although the hardness data indicate a correlation with morphology, the values are not quantitatively accurate. 

Unfortunately, the surface brightness of the torus and inner ring is so high that pile-up precludes accurate determination of the spectrum of these regions.
However, we did analyze the spectral distribution, using XSPEC (\cite{Arnaud1996}), of the trailed image and of the timed-exposure image of the outer Nebula.
For the trailed image, we fit the spectrum obtained from region T of Figure~\ref{f:image-full}, which is matched to the same range of columns covered by region Z (used in the hardness analysis), but excludes the Pulsar.  

Current uncertainties in the detector's response below 1~keV and in the zeroth-order effective area result in systematic errors in the spectral analysis.
Consequently, we are skeptical of the absolute values of the best-fit spectral parameters.
However, because these systematic effects comparably influence the fits for the trailed image and for the outer Nebula, we believe that the relative values are significant.
For a column density $N_{\rm H} = 3.45\! \times \! 10^{21} {\rm cm}\!^{-2}$ (\cite{Schattenburg1986}), we find that the spectral index of the outer Crab Nebula is steeper than that of the average Nebula by 0.62; for $2.35\! \times \! 10^{21} {\rm cm}\!^{-2}$ (formal best-fit value), by 0.50.

\newpage

\newpage

\newpage

\newpage

\end{document}